\documentclass{article}

\usepackage{arxiv}

\usepackage[utf8]{inputenc} % allow utf-8 input
\usepackage[T1]{fontenc}    % use 8-bit T1 fonts
\usepackage{hyperref}       % hyperlinks
\usepackage{url}            % simple URL typesetting
\usepackage{booktabs}       % professional-quality tables
\usepackage{amsfonts}       % blackboard math symbols
\usepackage{nicefrac}       % compact symbols for 1/2, etc.
\usepackage{microtype}      % microtypography
\usepackage{graphicx}
\usepackage[square,numbers]{natbib}
\usepackage{doi}

\title{ligandformer: a graph neural network for predicting compound property with robust  interpretation }

\author{ \href{https://orcid.org/0000-0002-9157-2199}{\includegraphics[scale=0.06]{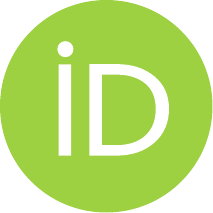}\hspace{1mm}Jinjiang Guo*}, {\hspace{2mm}Qi Liu}, {\hspace{2mm}Han Guo}, {\hspace{2mm}Xi Lu}\\
	\emph{Data Science group}\\
	Global Health Drug Discovery Institute, Beijing, China \\
	\texttt{jinjiang.guo@ghddi.org} \\
}

\date{}

\renewcommand{\headeright}{}
\renewcommand{\undertitle}{}

\hypersetup{
pdftitle={Ligandformer},
pdfsubject={cs.AI},
pdfauthor={Jinjiang GUO, Qi Liu, Han Guo, Xi Lu},
pdfkeywords={Graph Neural Networks, Compound Property Prediction, AI Model Interpretability, Attention Mechanism},
}

\begin{document}
\maketitle

\begin{abstract}
Robust and efficient interpretation of  QSAR methods  is quite useful to validate AI prediction rationales with subjective opinion (chemist or biologist expertise), understand sophisticated chemical or biological process mechanisms, and provide heuristic ideas for structure optimization in pharmaceutical industry. For this purpose, we construct a multi-layer self-attention  based Graph Neural Network framework, namely Ligandformer,  for predicting compound property with interpretation. Ligandformer  integrates  attention maps on compound structure from different network blocks. The integrated attention map reflects the machine's local interest on compound structure, and indicates the relationship between predicted compound property and its structure. This work mainly contributes to three aspects: 1. Ligandformer directly opens the black-box of deep learning methods, providing local prediction rationales on chemical structures. 2. Ligandformer  gives robust prediction in different experimental rounds, overcoming the ubiquitous prediction instability of deep learning methods. 3. Ligandformer can be generalized to predict different chemical or biological properties with high performance. Furthermore, Ligandformer can simultaneously output specific property score and visible attention map on structure, which can support researchers to investigate chemical or biological property and optimize structure efficiently. Our framework outperforms over counterparts in terms of accuracy, robustness and generalization, and can be applied in complex system study.
\end{abstract}

% keywords can be removed

\section{Introduction}
\label{sec:intro}
Recent advances of deep learning (DL) methods \cite{wu2018moleculenet, feinberg2019step, sarkar2016qsar, shao2015mining, wang2019molecule, liu2019chemi, goulon2007predicting,tang2020self} boost  the  performance of  quantitative structure-activity relationships (QSAR) models on predicting chemical or biological properties of molecules in drug discovery industry \cite{cherkasov2014qsar, chen2018rise}.  However, many DL models function as black-boxes \cite{goulon2007predicting}, which means that given a molecule with physiochemical/biological and structural features, DL models usually predict a simple global score  for certain desired property, leaving the inference rationales, such like local judgement on chemical structure, an unknown status.  Interpretation of such rationales  is useful  to  reveal relationship between structure and property, optimize compound structure,  and validate DL  models with subjective opinion (chemical or biological knowledge) \cite{cherkasov2014qsar, matveieva2021benchmarks, tang2020self}. Especially, recent Graph Neural Network (GNN) based methods \cite{wu2018moleculenet, liu2019chemi, hamilton2017inductive, xu2018powerful, morris2019weisfeiler, ranjan2020asap} were wildly designed  and  applied as QSAR models for predicting compound properties. Although GNN  methods outperform  against traditional machine learning methods such as random forest (RF) \cite{pal2005random} and support vector machines (SVM) \cite{noble2006support}, most of them still function as black-boxes, providing uninterpretable results \cite{goulon2007predicting}. In this work, we propose  a multi-layer self-attention  based Graph Neural Network framework, namely Ligandformer,  for predicting compound property with robust interpretation, which reflects machine interest on local region of input molecule.\footnote{Source code, pretrained-ready configurations, and the datasets used in this paper are publicly available at \url{https://github.com/GHDDI-AILab/LigandFormer}.}   

\section{Related Work}
\label{sec:relatedwork}
\subsection{Molecule Representation}
\label{sec:molrepresent}
For molecule chemical representation, current GNN methods  \cite{cherkasov2014qsar, matveieva2021benchmarks, tang2020self} usually process 2D graph as  description of natural chemical graph, in  which nodes  represent  atoms integrating different chemical attributes,  and edges represent  bonds  connecting atoms to one another. There are mainly three advantages of using 2D graph description: (1) graph  preserves clear and stable information of chemical structure, (2) it   represents invariant molecule regardless of entry position in line notation (e.g., SMILES \cite{weininger1988smiles}), (3) it can be easily computed and optimized by GNN methods.  In our chemical formulation, we take similar 2D graph representation, meanwhile we adopt bidirectional graph where the bond connection from atom A to atom B is the same as the bond connection from atom B and atom A. Moreover, 7 atomic chemical attributes, listed in Figure \ref{framework}, are considered as node initial features of input graph. Our GNN method learns and aggregates these attributes to be proper  molecular features for predicting certain property. For data preprocessing,  Ligandformer converts SMILES sequence into  2D graph formula, and node attributes in Figure \ref{framework} \textbf{c} can be used for distinguishing those compounds with same molecular structures. Hence, each molecule representation in our method is unique. 

\subsection{Self-Attention Mechanism}
\label{sec:selfattention}
Self-attention mechanism was firstly designed as key characteristic in Transformers \cite{vaswani2017attention, brown2020language, raffel2019exploring, parmar2018image, carion2020end, tay2020efficient, tang2020self}.  The mechanism can be considered as a graph-like inductive bias which relevantly pools features of  all tokens in an input sample. Such bias can be interpreted as machine's focus on each element of input sample.

The core operation of the mechanism is called single-head self-attention mechanism, in which each element learns to relevantly aggregate over all elements of an input $X$. The definition of a single-head self-attention mechanism is as follows:
\begin{equation} \label{eq1}
	Y_h = softmax(Q_h \cdot K^T_h)\cdot V_h
\end{equation}
where $Q_h = W_q X_h$, $K_h = W_k X_h$ and $T_h = W_vX_h$ are linear projections applied on the hidden features  $X_h \in \mathbb{R}^d$ of the input $X$. $W_q, W_k, W_v \in \mathbb{R}^{d \times d}$ are the weight matrices for query, key and value transformations, and $d$ is the dimension number of hidden input features $X_h$. In equation (\ref{eq1}), the attention matrix $A_h = Q_h \cdot K^T_h$ takes charge of learning relevant scores between any two elements of input $X_h$, which gives the reason why the process is called self-attention. Figure \ref{framework} \textbf{b} shows the working mechanism of a single-head self-attention layer.  For multi-head self-attention, the operations on $X_h$ are densely conducted $N$ times in parallel, which thereby output a series of ${Y_h}^i, i \in [1, 2, ..., N]$. The final output $Y_h = W_o[{Y_h}^1  {Y_h}^2  ... {Y_h}^N]$, where $W_o$ is the translation parameters for output linear projection.

\section{Method and Experiments}
\label{sec:method}
In this section, we describe our interpretable deep learning framework, namely Ligandformer, and demonstrate its performance on predicting compound's chemical/biological properties. For a given molecule, we utilize Python RDkit \cite{landrum2013rdkit} tool kit to process molecule SMILES as input data format, and convert the data into bidirectional graph $G$ based on Deepchem and Chemprop \cite{ramsundar2019deep, yang2019analyzing} processing ways. The graph mainly consists of index lists of nodes and edges. For each node, each attribute in Figure \ref{framework} \textbf{c} is converted into identical number, and all 7 numeric attributes are  combined as initial node feature vector $f_{init}$. The  node feature $f_{init}$ is then delivered to node embedding layer of our framework for learning optimal attribute combination, from which we can  obtain an aligned node feature vector $f_v$.

\subsection{Self-attention based Graph Neural Networks Architecture}
\begin{figure}
	\centering
	\includegraphics[scale=.34]{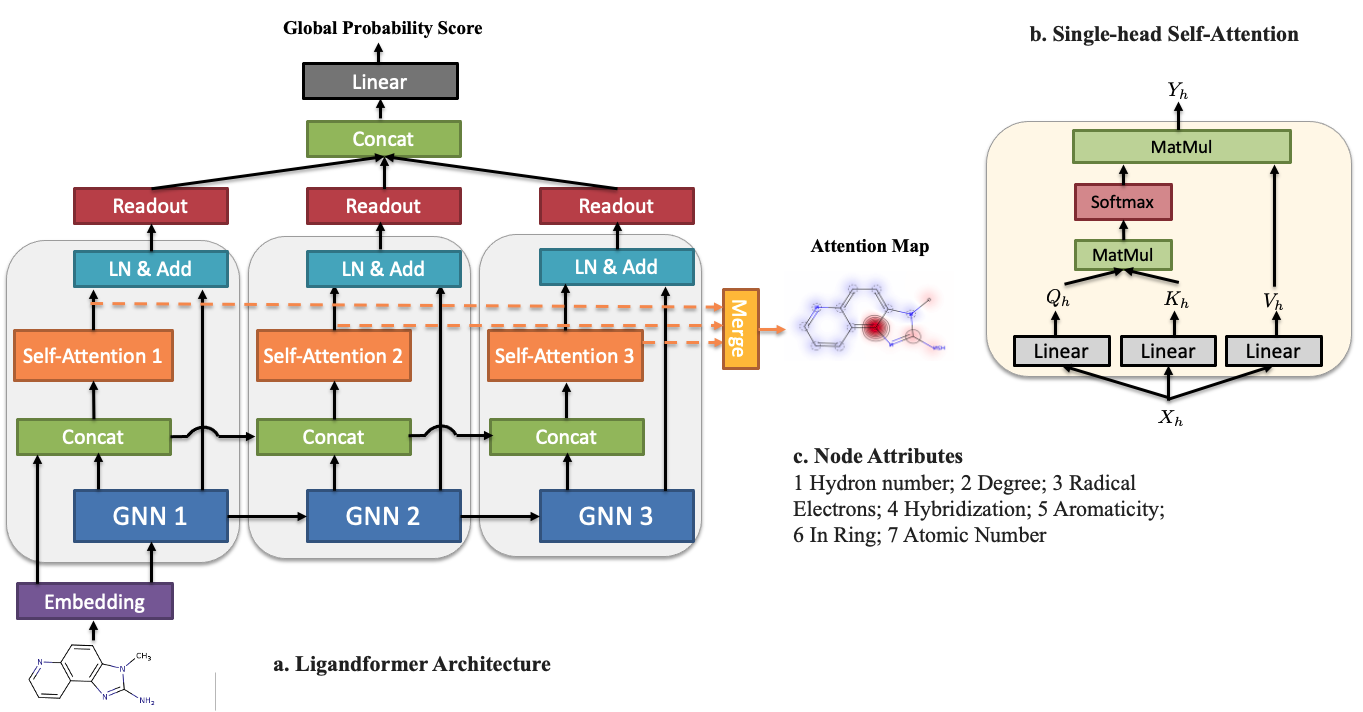}
	\caption{a. Ligandformer architecture; b. Single-head self-attention mechanism; c. Graph node initial attributes.}
	\label{framework}
\end{figure}

Similar to other Transformers, the block of our graph neural network framework is yet characterized by a graph neural network (GNN) module,  a multi-head  self-attention mechanism with configurable \texttt{num\_heads}, layer normalization layers \cite{ba2016layer} and residual passages. However, Ligandformer forms as a wide and parallel architecture, showed in Figure \ref{framework} \textbf{a}, and all hidden features from previous blocks are concatenated and fed to  self-attention layer of next block. The idea behind this dense connection is to enhance the message passing over different self-attention layers, and thus improve the robustness of attention mechanism. Different from SAMPN \cite{tang2020self}, another recent interpretable GNN method which concatenates self-attention mechanism between MPNN \cite{wu2018moleculenet} backbone and classifier/regressor, Ligandformer has a very different architecture that integrates self-attention mechanism into each computational block and merges attention information from different layers for final visible interpretation.  We describe  Ligandfomer architecture details and demonstrate its strengths  in the following parts of this paper.

For graph feature aggregation and message passing, we modified the structure of GIN \cite{xu2018powerful} and used it as our GNN module. The module is a variant of spatial based graph neural networks  \cite{velivckovic2017graph, kipf2016semi, hamilton2017inductive}, which aggregates features by taking both  summation and maxima of neighbouring features, and thus enhances the message propagation from shallow blocks to deep blocks.  Specifically, for each node $v$ in graph $G$, its node features $f_v^k$ from GNN module in $k$th block is calculated as:
\begin{equation}
\label{eq:hagnet}
f_v^k = \phi(\mathbf{concat}(f_v^{k-1}, (\sum_{u \in \mathcal{N}(v)}{f_u^{k-1}} + \max_{u \in \mathcal{N}(v)}{f_u^{k-1}} )))
\end{equation}
where $\phi(\cdot)$ is the MLP function \cite{gardner1998artificial}, and $\mathbf{concat(\cdot)}$ concatenates  features of node $v$ from $(k-1)$th  block and  $k$th block, $u$ denotes a neighbouring node in node $v$'s neighbourhood $\mathcal{N}(v)$. The output node feature $f_v^k$ are concatenated with node features  from all the previous blocks, i.e., $F_v^k = \mathbf{concat}(f_v^k, f_v^{k-1}, ...,  f_v^1 )$, and then feed $F_v^k$ to a multi-head self-attention layer (with configurable \texttt{num\_heads}) using equation (\ref{eq1}), the  final node feature $f_{vo}^k$ from $k$th block is computed as:
\begin{equation} \label{blockoutput}
	f_{vo}^k = LayerNorm(SelfAttention(F_v^k )) \oplus f_v^k
\end{equation}
Where $LayerNorm$ means layer normalization \cite{ba2016layer} and  $\oplus$ is the element wise addition along the feature vector. We then take a $\mathbf{mean}$ pooling \emph{read-out} function over all the nodes of graph $G$,  and thus obtain its graph representation $f_G^k$  from $k$th block.  The graph representations from all the layers are concatenated  and fed to  a 3-layer MLP $\Phi(\cdot)$  functional head to predict property probability $\mathbf{p}_G$:
\begin{equation}
\label{eq:mlphead}
\mathbf{p}_G = \Phi(\mathbf{concat}(f_G^1, f_G^2, ..., f_G^k, ...,  f_G^K))
\end{equation}
in which $K$ is the block number.

For loss function and optimization,  in this work we mainly investigate the binary classification problem, i.e., positive or negative of certain property, and thus we adopt cross-entropy as our loss function, which can be simply written as:
\begin{equation}
\label{eq:loss}
\mathcal{L_G} =-\log{\mathbf{p}_G}
\end{equation}
Also, our method can be applied in different kinds of tasks, e.g,. regression, multi-classification, ranking, etc., by modifying the corresponding functional head (e.g., regressor, classifier, etc.) and loss function. We tried SGD \cite{kiwiel2001convergence}, Adam \cite{kingma2014adam} and Adabelief \cite{zhuang2020adabelief} to optimize model parameters, and find that Adam provides the best optimal model with highest evaluation performance and most stable convergence during training and testing.
 
\subsection{Dataset and Experiments}
To verify the generalization ability of Ligandformer, we selected three chemical properties  from  different domains in drug research \& development pipelines:  1. aqueous solubility, 2. Caco-2 cell permeabilty and 3. Ames mutagenesis from ADME/T domains. We trained and tested Ligandformer on public disclosed datasets of these three properties.
\subparagraph{Aqueous Solubility} Aqueous solubility is the chemical saturated concentration in the aqueous phase, which is usually described with unit log(mol/L) and written as logS. This dataset was downloaded from the website, namely online chemical database and modeling environment (OCHEM) \cite{sushko2011online}, which includes 1,311 assay records. To ensure this task is binary classification, we set  $LogS=0$ as  cut-off for distinguishing soluble or insoluble of each compounds in the dataset.
\subparagraph{Caco-2 Cell Permeabilty}  Caco-2 cell permeability \cite{van2005caco} assay measures the rate of flux of a chemical across polarised Caco-2 cell monolayers and the results from the assay  can be used to predict drug \emph{in vivo}  absorption. We collected and cleaned 7,624 assayed data from public and commercial sources, e.g., REAXYS \cite{currano2014chemical}, and set  $Papp \geq 10 \times 10^{-6}cm/s$ as a threshold for judging positive and negative  w.r.t absorption.
\subparagraph{Ames Mutagenesis} Ames mutagenesis \cite{mortelmans2000ames} assay usually uses bacteria to test whether a given chemical can cause mutations in the DNA of the test organism. A positive assayed result  indicates that the chemical is mutagenic, which implies high risk of inducing cancer. In our case,  7,617 assayed data were collected from public sources as our training and testing dataset. We set  Colonies $=$ 2 fold  as the cut-off for binary classification, where $Colonies \geq 2 fold$ means positive, otherwise negative. 

We trained and tuned each model’s parameters under same default protocol with three properties respectively. In details, we took a 90-10 split on three datasets  respectively, where 90\% of dataset was used for training and the remaining 10\% was used for testing.  Random shuffle on each dataset was used for ensuring that each input data sample give an independent change on the model in each training batch. Also, we resmapled the positive and negative data of training samples, balancing their number ratio to 1:1, which avoids the training skewness towards the class with majority samples. For the hyperparameters configuration, we configured 3 characterized blocks for the architecture, and selected Adam optimizer with learning rate $\eta = 0.001$ and weight decay $ \lambda = 0.0001$. The training batch size was fixed to 256, and we utilized early stop = 50 to avoid overfitting, which means the training procedure will stop after 50 epochs when the testing performance (i.e., AUROC) does not continue to increase at all.  We compared Ligandformer with recent designed  counterparts, i.e., MPNN, SAMPN  w.r.t peformance on predicting three chemical properties. As we can see from Table \ref{tab:performance}, Ligandformer constantly outperforms in terms of AUROC, reaching up to 0.98, 0.89 and 0.92 respectively. Furthermore, the self-attention module in Ligandformer exposes a configurable \texttt{num\_heads} hyperparameter (set to 1 by default in our experiments), and we believe that increasing \texttt{num\_heads} can further leverage Ligandformer's performance on predicting  chemical/biological properties.

\begin{table}
	\caption{AUC performance comparison on  aqueous solubility, Caco-2 cell permeabilty and  Ames mutagenesis.}
	\centering
	\begin{tabular}{cccc}
		\toprule
		     &  Aqueous Solubility     & Caco-2 Cell Permeabilty 	&Ames Mutagenesis  \\
		\midrule
		MPNN		 & 0.93  			& \textbf{0.89}			&0.90	\\
		SAMPN		 & 0.92				& 0.88    		&0.91\\
		Ligandformer (ours) & \textbf{0.98}  & \textbf{0.89} &\textbf{0.92}\\
		\bottomrule
	\end{tabular}
	\label{tab:performance}
\end{table}

\subsection{Attention Map Visualization}

\begin{figure}
	\centering
	\includegraphics[scale=.30]{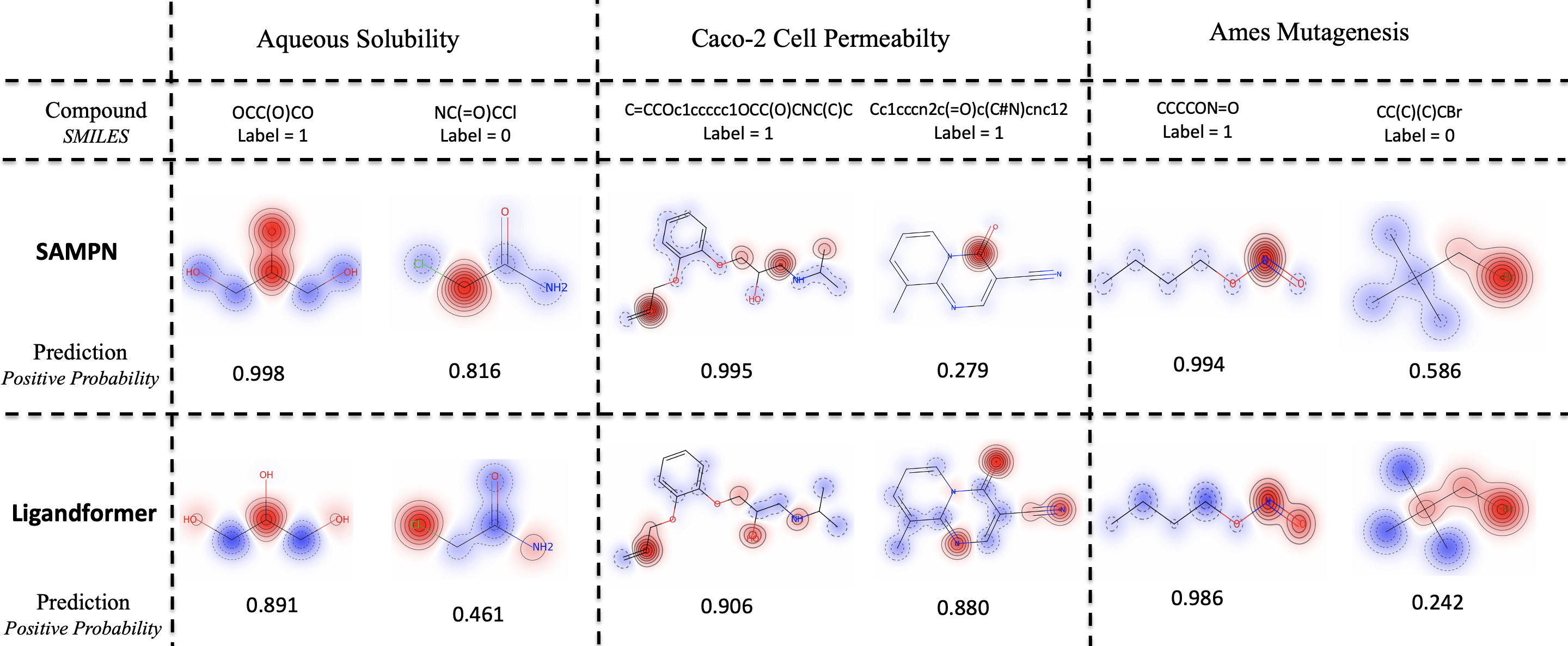}
	\caption{Attention maps generated by SAMPN and Ligandformer for three chemical properties. In each heat map, redder color indicates predicted auxo-action feature on specific property.}
	\label{att_viz}
\end{figure}

\begin{figure}
	\centering
	\includegraphics[scale=.30]{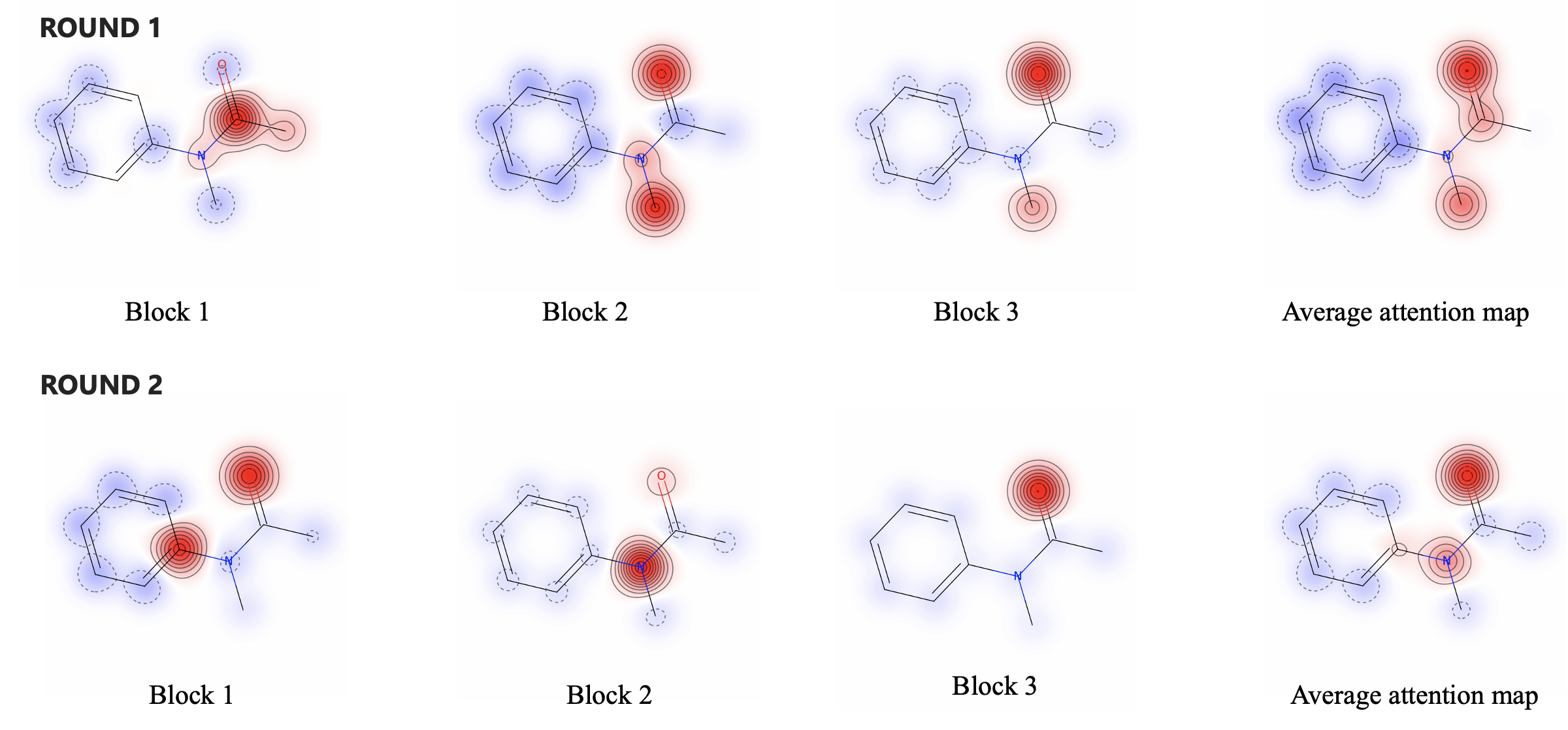}
	\caption{Even though two corresponding attention maps of the same block are different in round 1 and round 2, the final  corresponding integrated (averaged) attention maps remain the same in general.}
	\label{diff_rounds}
\end{figure}

Throughout the discussion before, while perusing higher prediction accuracy, robust and efficient interpretation on QSAR model is also important. By utilizing visualization technique on attention mechanism, researchers can directly validate and identify  the learned features that determine compound property predictions. In Ligandformer, for an input molecule, each block $i$ can provide relevant attention score matrix $\mathcal{M}_i^{n \times n}$ for all $n$ atoms. In $\mathcal{M}_i^{n \times n}$, while the summation of each column equals to 1, we define the average value of each row of  $\mathcal{M}_i^{n \times n}$ as  attention coefficients to observe each atom's contribution to the property.  While attention coefficients from shallow blocks represents  atomic or local parts contribution, attention coefficients from deeper blocks reflects fragments or larger scale parts contribution. We calculate the average of attention coefficients from all blocks to gain an integrated attention map from different scopes. Using heat map to visualize the integrated attention helps chemist/biologist to find which parts of molecule play a more important role in certain property, and therefore assist researchers to optimize compound structure accordingly. Figure \ref{att_viz} shows several visualization examples given by Ligandformer and SAMPN respectively. 
\subparagraph{Robustness validation} To our knowledge, deep learning models often suffer from prediction instability. Even though the overall testing performance stays the same in different training rounds, specific prediction scores for the same  testing sample are usually different. This is mainly because of the random initial model parameters. Ligandformer overcomes such instability by integrating all the attention coefficients from different blocks.  To demonstrate  the robustness of attention map, we launched two different training rounds with random initial model parameters on same training and testing datasets of aqueous solubility. As shown in Figure \ref{diff_rounds}, even though two corresponding attention maps of the same block are different in different training rounds, the final  corresponding integrated (averaged) attention maps remain the same in general. It indicates that, given same training materials on certain property,  Ligandformer can output robust attention map.

\section{Conclusion}
In this work, we have introduced our Ligandformer,  a multi-layer multi-head self-attention  based Graph Neural Network framework (with configurable \texttt{num\_heads}),  for predicting compound chemical property with robust interpretation, i.e.,  integrated attention map.  Facilitated with visualization technique, the map shows  insights on AI model's rationales on judging which parts of an input molecule  impact certain property predictions, and thus it can support researchers to investigate chemical or biological property and optimize structure efficiently.  Our Ligandformer outperforms over recent GNN counterparts, giving more robust interpretation.  Ligandformer also demonstrated a good generalization ability  to predict different chemical properties with high accuracy. We hope this interpretable QSAR method can contribute to scientific community and drug discovery industry as well.

\section*{Code and Data Availability}
The full Ligandformer implementation, configuration files, training/evaluation/prediction scripts, the pre-processed Aqueous Solubility, Caco-2 Cell Permeability, and Ames Mutagenesis datasets used in this paper, and the \LaTeX{} source of this manuscript are released at \url{https://github.com/GHDDI-AILab/LigandFormer}.

\bibliographystyle{unsrtnat}
\bibliography{references}

@article{kipf2016semi,
  title={Semi-supervised classification with graph convolutional networks},
  author={Kipf, Thomas N and Welling, Max},
  journal={arXiv preprint arXiv:1609.02907},
  year={2016}
}

@article{hamilton2017inductive,
  title={Inductive representation learning on large graphs},
  author={Hamilton, William L and Ying, Rex and Leskovec, Jure},
  journal={arXiv preprint arXiv:1706.02216},
  year={2017}
}

@article{xu2018powerful,
  title={How powerful are graph neural networks?},
  author={Xu, Keyulu and Hu, Weihua and Leskovec, Jure and Jegelka, Stefanie},
  journal={arXiv preprint arXiv:1810.00826},
  year={2018}
}

@inproceedings{ranjan2020asap,
  title={Asap: Adaptive structure aware pooling for learning hierarchical graph representations},
  author={Ranjan, Ekagra and Sanyal, Soumya and Talukdar, Partha},
  booktitle={Proceedings of the AAAI Conference on Artificial Intelligence},
  volume={34},
  number={04},
  pages={5470--5477},
  year={2020}
}

@article{kiwiel2001convergence,
  title={Convergence and efficiency of subgradient methods for quasiconvex minimization},
  author={Kiwiel, Krzysztof C},
  journal={Mathematical programming},
  volume={90},
  number={1},
  pages={1--25},
  year={2001},
  publisher={Springer}
}

@article{kingma2014adam,
  title={Adam: A method for stochastic optimization},
  author={Kingma, Diederik P and Ba, Jimmy},
  journal={arXiv preprint arXiv:1412.6980},
  year={2014}
}

@article{zhuang2020adabelief,
  title={Adabelief optimizer: Adapting stepsizes by the belief in observed gradients},
  author={Zhuang, Juntang and Tang, Tommy and Tatikonda, Sekhar and Dvornek, Nicha and Ding, Yifan and Papademetris, Xenophon and Duncan, James S},
  journal={arXiv preprint arXiv:2010.07468},
  year={2020}
}

@article{wu2018moleculenet,
  title={MoleculeNet: a benchmark for molecular machine learning},
  author={Wu, Zhenqin and Ramsundar, Bharath and Feinberg, Evan N and Gomes, Joseph and Geniesse, Caleb and Pappu, Aneesh S and Leswing, Karl and Pande, Vijay},
  journal={Chemical science},
  volume={9},
  number={2},
  pages={513--530},
  year={2018},
  publisher={Royal Society of Chemistry}
}

@misc{feinberg2019step,
  title={Step Change Improvement in ADMET Prediction with PotentialNet Deep Featurization. arXiv. org},
  author={Feinberg, EN and Sheridan, R and Joshi, E and Pande, VS and Cheng, AC},
  year={2019},
  publisher={March}
}

@article{sarkar2016qsar,
  title={QSAR STUDIES OF FabH INHIBITORS USING GRAPH THEORETICAL \& QUANTUM CHEMICAL DESCRIPTORS.},
  author={Sarkar, Dipanjan and Sharma, Shyamal and Mukhopadhyay, Subhasis and Bothra, Asim Kumar},
  journal={Pharmacophore},
  volume={7},
  number={4},
  year={2016}
}

@article{shao2015mining,
  title={Mining discriminative patterns from graph data with multiple labels and its application to quantitative structure--activity relationship (Qsar) models},
  author={Shao, Zheng and Hirayama, Yuya and Yamanishi, Yoshihiro and Saigo, Hiroto},
  journal={Journal of chemical information and modeling},
  volume={55},
  number={12},
  pages={2519--2527},
  year={2015},
  publisher={ACS Publications}
}

@article{wang2019molecule,
  title={Molecule property prediction based on spatial graph embedding},
  author={Wang, Xiaofeng and Li, Zhen and Jiang, Mingjian and Wang, Shuang and Zhang, Shugang and Wei, Zhiqiang},
  journal={Journal of chemical information and modeling},
  volume={59},
  number={9},
  pages={3817--3828},
  year={2019},
  publisher={ACS Publications}
}

@article{liu2019chemi,
  title={Chemi-Net: a molecular graph convolutional network for accurate drug property prediction},
  author={Liu, Ke and Sun, Xiangyan and Jia, Lei and Ma, Jun and Xing, Haoming and Wu, Junqiu and Gao, Hua and Sun, Yax and Boulnois, Florian and Fan, Jie},
  journal={International journal of molecular sciences},
  volume={20},
  number={14},
  pages={3389},
  year={2019},
  publisher={Multidisciplinary Digital Publishing Institute}
}

@article{goulon2007predicting,
  title={Predicting activities without computing descriptors: graph machines for QSAR},
  author={Goulon, A and Picot, T and Duprat, A and Dreyfus, G},
  journal={SAR and QSAR in Environmental Research},
  volume={18},
  number={1-2},
  pages={141--153},
  year={2007},
  publisher={Taylor \& Francis}
}

@article{tang2020self,
  title={A self-attention based message passing neural network for predicting molecular lipophilicity and aqueous solubility},
  author={Tang, Bowen and Kramer, Skyler T and Fang, Meijuan and Qiu, Yingkun and Wu, Zhen and Xu, Dong},
  journal={Journal of cheminformatics},
  volume={12},
  number={1},
  pages={1--9},
  year={2020},
  publisher={BioMed Central}
}

@article{cherkasov2014qsar,
  title={QSAR modeling: where have you been? Where are you going to?},
  author={Cherkasov, Artem and Muratov, Eugene N and Fourches, Denis and Varnek, Alexandre and Baskin, Igor I and Cronin, Mark and Dearden, John and Gramatica, Paola and Martin, Yvonne C and Todeschini, Roberto and others},
  journal={Journal of medicinal chemistry},
  volume={57},
  number={12},
  pages={4977--5010},
  year={2014},
  publisher={ACS Publications}
}

@inproceedings{morris2019weisfeiler,
  title={Weisfeiler and leman go neural: Higher-order graph neural networks},
  author={Morris, Christopher and Ritzert, Martin and Fey, Matthias and Hamilton, William L and Lenssen, Jan Eric and Rattan, Gaurav and Grohe, Martin},
  booktitle={Proceedings of the AAAI Conference on Artificial Intelligence},
  volume={33},
  number={01},
  pages={4602--4609},
  year={2019}
}

@article{matveieva2021benchmarks,
  title={Benchmarks for interpretation of QSAR models},
  author={Matveieva, Mariia and Polishchuk, Pavel},
  journal={Journal of cheminformatics},
  volume={13},
  number={1},
  pages={1--20},
  year={2021},
  publisher={Springer}
}

@article{vaswani2017attention,
  title={Attention is all you need},
  author={Vaswani, Ashish and Shazeer, Noam and Parmar, Niki and Uszkoreit, Jakob and Jones, Llion and Gomez, Aidan N and Kaiser, {\L}ukasz and Polosukhin, Illia},
  journal={Advances in neural information processing systems},
  volume={30},
  year={2017}
}

@article{brown2020language,
  title={Language models are few-shot learners},
  author={Brown, Tom and Mann, Benjamin and Ryder, Nick and Subbiah, Melanie and Kaplan, Jared D and Dhariwal, Prafulla and Neelakantan, Arvind and Shyam, Pranav and Sastry, Girish and Askell, Amanda and others},
  journal={Advances in neural information processing systems},
  volume={33},
  pages={1877--1901},
  year={2020}
}

@article{raffel2019exploring,
  title={Exploring the limits of transfer learning with a unified text-to-text transformer},
  author={Raffel, Colin and Shazeer, Noam and Roberts, Adam and Lee, Katherine and Narang, Sharan and Matena, Michael and Zhou, Yanqi and Li, Wei and Liu, Peter J},
  journal={arXiv preprint arXiv:1910.10683},
  year={2019}
}

@inproceedings{parmar2018image,
  title={Image transformer},
  author={Parmar, Niki and Vaswani, Ashish and Uszkoreit, Jakob and Kaiser, Lukasz and Shazeer, Noam and Ku, Alexander and Tran, Dustin},
  booktitle={International Conference on Machine Learning},
  pages={4055--4064},
  year={2018},
  organization={PMLR}
}

@inproceedings{carion2020end,
  title={End-to-end object detection with transformers},
  author={Carion, Nicolas and Massa, Francisco and Synnaeve, Gabriel and Usunier, Nicolas and Kirillov, Alexander and Zagoruyko, Sergey},
  booktitle={European conference on computer vision},
  pages={213--229},
  year={2020},
  organization={Springer}
}

@article{tay2020efficient,
  title={Efficient transformers: A survey},
  author={Tay, Yi and Dehghani, Mostafa and Bahri, Dara and Metzler, Donald},
  journal={arXiv preprint arXiv:2009.06732},
  year={2020}
}

@article{chen2018rise,
  title={The rise of deep learning in drug discovery},
  author={Chen, Hongming and Engkvist, Ola and Wang, Yinhai and Olivecrona, Marcus and Blaschke, Thomas},
  journal={Drug discovery today},
  volume={23},
  number={6},
  pages={1241--1250},
  year={2018},
  publisher={Elsevier}
}

@article{weininger1988smiles,
  title={SMILES, a chemical language and information system. 1. Introduction to methodology and encoding rules},
  author={Weininger, David},
  journal={Journal of chemical information and computer sciences},
  volume={28},
  number={1},
  pages={31--36},
  year={1988},
  publisher={ACS Publications}
}

@book{ramsundar2019deep,
  title={Deep learning for the life sciences: applying deep learning to genomics, microscopy, drug discovery, and more},
  author={Ramsundar, Bharath and Eastman, Peter and Walters, Patrick and Pande, Vijay},
  year={2019},
  publisher={O'Reilly Media}
}

@article{yang2019analyzing,
  title={Analyzing learned molecular representations for property prediction},
  author={Yang, Kevin and Swanson, Kyle and Jin, Wengong and Coley, Connor and Eiden, Philipp and Gao, Hua and Guzman-Perez, Angel and Hopper, Timothy and Kelley, Brian and Mathea, Miriam and others},
  journal={Journal of chemical information and modeling},
  volume={59},
  number={8},
  pages={3370--3388},
  year={2019},
  publisher={ACS Publications}
}

@misc{landrum2013rdkit,
  title={RDKit: A software suite for cheminformatics, computational chemistry, and predictive modeling},
  author={Landrum, Greg and others},
  year={2013},
  publisher={Academic Press Cambridge}
}

@article{ba2016layer,
  title={Layer normalization},
  author={Ba, Jimmy Lei and Kiros, Jamie Ryan and Hinton, Geoffrey E},
  journal={arXiv preprint arXiv:1607.06450},
  year={2016}
}

@article{velivckovic2017graph,
  title={Graph attention networks},
  author={Veli{\v{c}}kovi{\'c}, Petar and Cucurull, Guillem and Casanova, Arantxa and Romero, Adriana and Lio, Pietro and Bengio, Yoshua},
  journal={arXiv preprint arXiv:1710.10903},
  year={2017}
}

@article{sushko2011online,
  title={Online chemical modeling environment (OCHEM): web platform for data storage, model development and publishing of chemical information},
  author={Sushko, Iurii and Novotarskyi, Sergii and K{\"o}rner, Robert and Pandey, Anil Kumar and Rupp, Matthias and Teetz, Wolfram and Brandmaier, Stefan and Abdelaziz, Ahmed and Prokopenko, Volodymyr V and Tanchuk, Vsevolod Y and others},
  journal={Journal of computer-aided molecular design},
  volume={25},
  number={6},
  pages={533--554},
  year={2011},
  publisher={Springer}
}

@article{van2005caco,
  title={Caco-2 cell permeability assays to measure drug absorption},
  author={van Breemen, Richard B and Li, Yongmei},
  journal={Expert opinion on drug metabolism \& toxicology},
  volume={1},
  number={2},
  pages={175--185},
  year={2005},
  publisher={Taylor \& Francis}
}

@article{mortelmans2000ames,
  title={The Ames Salmonella/microsome mutagenicity assay},
  author={Mortelmans, Kristien and Zeiger, Errol},
  journal={Mutation research/fundamental and molecular mechanisms of mutagenesis},
  volume={455},
  number={1-2},
  pages={29--60},
  year={2000},
  publisher={Elsevier}
}

@book{currano2014chemical,
  title={Chemical information for chemists: a primer},
  author={Currano, Judith and Roth, Dana},
  year={2014},
  publisher={Royal Society of Chemistry}
}

@article{gardner1998artificial,
  title={Artificial neural networks (the multilayer perceptron)—a review of applications in the atmospheric sciences},
  author={Gardner, Matt W and Dorling, SR},
  journal={Atmospheric environment},
  volume={32},
  number={14-15},
  pages={2627--2636},
  year={1998},
  publisher={Elsevier}
}

@article{pal2005random,
  title={Random forest classifier for remote sensing classification},
  author={Pal, Mahesh},
  journal={International journal of remote sensing},
  volume={26},
  number={1},
  pages={217--222},
  year={2005},
  publisher={Taylor \& Francis}
}

@article{noble2006support,
  title={What is a support vector machine?},
  author={Noble, William S},
  journal={Nature biotechnology},
  volume={24},
  number={12},
  pages={1565--1567},
  year={2006},
  publisher={Nature Publishing Group}
}

\end{document}